\documentstyle[12pt,epsf]{ioplppt}

\begin{document}

\jl{1}

\title{The $A+B \rightarrow \emptyset$ annihilation
reaction in a quenched random velocity field}[The $A+B
\rightarrow \emptyset$ reaction in a random velocity field]

\author{K Oerding\ftnote{1}{Address after 1 December 1996:
Institut f\"ur Theoretische Physik III,
Heinrich-Heine-Universit\"at, D-40225 D\"usseldorf,
Germany}
}

\address{Department of Physics, Theoretical Physics,
1 Keble Road, Oxford OX1 3NP, UK}

\begin{abstract}
Using field-theoretic renormalization group methods the long-time
behaviour of the $A+B \rightarrow \emptyset$ annihilation
reaction with equal initial densities
$n_{\rm A}(0) = n_{\rm B}(0) = n_{0}$ in a quenched random
velocity field is studied. At every point $(x, {\bf y})$
of a $d$-dimensional system the velocity is parallel or
antiparallel to the $x$-axis and depends on the coordinates
perpendicular to the flow. Assuming that $v({\bf y})$ have zero
mean and short-range correlations in the y-direction we show that
the densities decay asymptotically as
$n_{\rm A, B}(t) \simeq A n_{0}^{1/2} t^{-(d+3)/8}$ for $d<3$.
The amplitude $A$ is calculated at first order in $\epsilon =
3-d$.
\end{abstract}

\maketitle

\section{Introduction}

It is well known that the time development of reaction diffusion
systems in which two types of particles A and B (`particles' and
`antiparticles') annihilate irreversibly is dominated by
density fluctuations~\cite{TW,KR,BL,LC1,SSB}.
If an equal number of A and B particles is randomly placed on
the sites of a $d$-dimensional lattice the densities are
asmptotically given by $n(t) \sim t^{-d/4}$ (for $d<4$) whereas a
na\"{\ii}ve approach based on mean-field rate equations predicts
$n(t) \sim t^{-1}$.

The slow density decay is due to the asymptotic segregation
of particles into regions of purely A or B particles. The reaction
becomes more efficient if the effects of segregation are reduced
by an appropriate mixing mechanism like turbulence~\cite{DW} or
random forces in low-viscosity liquids~\cite{AHLS}. In these
cases the diffusive particle motion is replaced by a
superdiffusive behaviour characterized by a mean-square
displacement of the form $\langle x(t)^{2}\rangle \sim t^{2/z}$
with $z < 2$. Very recently the annihilation reaction in
a driven diffusive system with repulsive interaction between
particles of the same type~\cite{Jan,IKR} has been studied. If
the particles are subject to an external driving force in one
direction the repulsion gives rise to superdiffusion with $z=3/2$
(for $d=1$), and the density decreases asymptotically as $n(t)
\sim t^{-1/3}$. The relation between the long-time behaviour of
$n(t)$ and the value of the exponent $z$ has been investigated
for particles performing independent L\'evy walks~\cite{ZK}. In
this case $z$ is defined through the scaling of the transition
probability (which is a function of $r^{-z} t$) since the the
mean-square displacement for L\'evy flights with $z<2$ diverges.
It has been shown that below the upper critical dimension
$d_{c}(z) = 2z$ the long-time behaviour of the density is
given by $n(t) \sim t^{-d/(2z)}$.

The mixing caused by the inhomogeneous velocity field in a linear
shear flow reduces the critical dimension to $d_{c}=2$, i.e. $n(t)
\sim t^{-1}$ in any physically accessible dimension~\cite{HB}.
Other models for particle mixing have been considered in 
references~\cite{SB1,SB2}. In this paper we investigate the
A+B-annihilation reaction in a medium with a random velocity
field. It is assumed that the velocity at every point ${\bf r}
= (x, {\bf y})$ of a $d$-dimensional system is parallel or
antiparallel to the $x$-axis and depends only on the coordinates
perpendicular to the flow. The velocity field is modelled by
quenched Gaussian random variables with zero mean and the
correlations
\begin{equation}
\left[ v({\bf y}) v({\bf y}^{\prime}) \right] = f \delta({\bf y}
- {\bf y}^{\prime}) \label{eq1}
\end{equation}
where the square brackets indicate the average over the
realizations of the flow. Originally this model was
introduced to describe the ground water transport in
heterogeneous rocks~\cite{MdeM}.

Below three dimensions a random walk in the presence of a
velocity field of the form~(\ref{eq1}) displays superdiffusive
behaviour in the $x$-direction~\cite{JH,BGKPR}. For a particle
starting at ${\bf r} = {\bf 0}$ at time $t=0$ the mean square
displacement in the $x$-direction averaged over the configurations
of $v({\bf y})$ is given by $[\langle x^{2} \rangle] \simeq
\sigma^{2} t^{(5-d)/2}$ for $d<3$ (with a generalized diffusion
constant $\sigma^{2}$). One can use this result for a na\"{\ii}ve
estimate of the densities at large times $t$. If we approximate
the motion of the particles by independent anisotropic L\'evy
flights with exponents $z_{\|} = 4/(5-d)$ (in $x$-direction) and
$z_{\bot}=2$ a straightforward generalization of the arguments
given in~\cite{ZK} yields
\begin{equation}
n_{\rm A, B}(t) \sim \left[ n_{0} t^{-1/z_{\|}}
t^{-(d-1)/z_{\bot}} \right]^{1/2} = n_{0}^{1/2} t^{-(d+3)/8}
\label{levy}
\end{equation}
where $n_{0}$ is the initial density of each particle type.
However, it is not clear in how far the effects of the velocity
field can be described by independent L\'evy walks since
$v({\bf y})$ is a {\em quenched} random variable with an
infinite correlation length in the $x$-direction.
In the following sections we will use the field theoretic
methods developed in~\cite{Pel,LC1} to show that the power
law~(\ref{levy}) is correct for $t \rightarrow \infty$ and
calculate the amplitude at first order in $\epsilon
= 3-d$. The result of this calculation reads
\begin{equation}
n_{\rm A, B}(t) \simeq A(\epsilon) \left( 2 \pi^{3/2} (8 \pi
D)^{(d-1)/2} \sigma \right)^{-1/2} \sqrt{n_{0}} t^{-(d+3)/8} 
\end{equation}
where
\begin{equation}
A(\epsilon) = 1 + \frac{\epsilon}{8} \left( 3 \ln 2 - 1 \right)
+ \Or(\epsilon^{2})
\end{equation}
and $n_{0} = n_{\rm A}(0)=n_{\rm B}(0)$ is the initial
density of each particle type.
In contrast to the case of annihilating L\'evy particles
which can be treated in a mean field approximation with random
initial conditions the correlated particle motion in a
quenched velocity field requires the application of the
renormalization group.

In the next section a microscopic model for a reaction diffusion
system in a random velocity field is defined and its mapping to a
continuum field theory is discussed. In Section~\ref{nonint}
we briefly review the renormalized field theory for
non-interacting particles in a random velocity field.
The results are used in Section~\ref{Seff} to derive an
effective action which describes the long-time behaviour
of the system. The asymptotic decay of the densities is
calculated in section~\ref{decay} and the results are discussed
in Section~\ref{concl}.

\section{The model} \label{model}

At every time $t$ the microscopic state of the system is defined
by the occupation numbers $m({\bf r}_{i})$ of A-particles and
$n({\bf r}_{i})$ of B-particles at every site $i$ (with position
${\bf r}_{i}$) of a $d$-dimensional lattice.
Denoting the probability for a given configuration $\{m, n \}$ at
time $t$ by $P(\{m, n\}; t)$ we can describe the stochastic
dynamics of the system by a set of linear differential equations
(master equation) for the probabilities $P(\{m, n\}; t)$.
(We assume that the time evolution is Markoffian.)
In order to apply field theoretic renormalization group methods
we use a functional integral description which is equivalent to
the master equation. Since a detailed derivation of this
formalism is given in references~\cite{Pel,BLee,LC1} we only give
the main steps focussing on the modifications that are necessary
to study the influence of the random velocity field.

We first map the probabilities $P(\{m, n\}; t)$ to a vector
$|\Phi(t)\rangle$ of the infinite dimensional Fock
space spanned by the vectors $|\{m, n\}\rangle$ by
writing~\cite{Doi,PG}
\begin{equation}
|\Phi(t)\rangle = \sum_{\{m, n\}} P(\{m, n\}; t)
\,|\{m, n\}\rangle .
\end{equation}
Since the master equation is linear and of first order in $t$ it
may be written in the form
\begin{equation}
\partial_{t} |\Phi(t)\rangle = \hat{L} |\Phi(t)\rangle
\label{Liou}
\end{equation}
with an appropriate Liouville operator $\hat{L}$. This operator
can be expressed in terms of the annihilation operators
$\hat{a}_{\rm A}({\bf r}_{i})$, $\hat{a}_{\rm B}({\bf r}_{i})$
and creation operators $\hat{a}_{\rm A}^{\star}({\bf r}_{i})$,
$\hat{a}_{\rm B}^{\star}({\bf r}_{i})$ defined by
\begin{equation}
\eqalign{
\hat{a}_{\rm A}({\bf r}_{i}) |m({\bf r}_{i}) \rangle = m({\bf
r}_{i}) |m({\bf r}_{i}) - 1 \rangle \qquad & \hat{a}_{\rm
A}^{\star}({\bf r}_{i}) |m({\bf  r}_{i}) \rangle = |m({\bf
r}_{i}) + 1 \rangle \\
\hat{a}_{\rm B}({\bf r}_{i}) |n({\bf r}_{i}) \rangle = n({\bf
r}_{i}) |n({\bf r}_{i}) - 1 \rangle & \hat{a}_{\rm
B}^{\star}({\bf r}_{i}) |n({\bf  r}_{i}) \rangle = |n({\bf
r}_{i}) + 1 \rangle .
}
\end{equation}
These operators leave the occupation numbers $m({\bf
r}_{j})$ and $n({\bf r}_{j})$ of the sites $j \neq i$ unchanged.

The diffusion of non-interacting particles corresponds to the
Liouvillean
\begin{eqnarray}
\fl \hat{L}_{D} = \frac{D}{h^{2}} \sum_{<i,j>} \left(
\hat{a}_{\rm A}^{\star}({\bf r}_{j}) - \hat{a}_{\rm
A}^{\star}({\bf r}_{i}) \right) \left( \hat{a}_{\rm A}({\bf
r}_{i}) - \hat{a}_{\rm A}({\bf r}_{j}) \right) \nonumber \\
+ \frac{D}{h^{2}} \sum_{<i,j>} \left( \hat{a}_{\rm
B}^{\star}({\bf r}_{j}) - \hat{a}_{\rm B}^{\star}({\bf r}_{i})
\right) \left( \hat{a}_{\rm B}({\bf r}_{i}) - \hat{a}_{\rm
B}({\bf r}_{j}) \right)
\end{eqnarray}
where the sum extends over all pairs of nearest neighbour sites,
$h$ is the lattice spacing and $D$ denotes the
diffusion constant. The annihilation of particle-antiparticle
pairs at the same site is described by
\begin{equation}
\hat{L}_{\rm reac} = \frac{k}{h^{d}} \sum_{i} \left( 1 -
\hat{a}_{\rm A}^{\star}({\bf r}_{i}) \hat{a}_{\rm
B}^{\star}({\bf r}_{i}) \right) \hat{a}_{\rm A}({\bf r}_{i})
\hat{a}_{\rm B}({\bf r}_{i})
\end{equation}
where $k/h^{d}$ is the reaction rate.

To investigate the effect of a random shear flow we introduce
independent Gaussian random variables $v({\bf y})$ labelled by
the coordinates ${\bf y}$ perpendicular to the direction of the
flow. At each point ${\bf r}_{i} = (x_{i}, {\bf y}_{i})$ the
velocity field prefers particle jumps parallel or antiparallel to
the $x$-direction depending on the sign of $v({\bf y}_{i})$. If
$v({\bf y}_{i})$ is positive particles at the point $(x_{i}, {\bf
y}_{i})$ will jump with rate $v({\bf y}_{i})/h$ in positive
$x$-direction whereas for $v({\bf y}_{i}) < 0$ they will hop with
rate $-v({\bf y}_{i})/h$ in the opposite direction . Here we
assume that $v({\bf y})$ have  zero mean and the correlations
\begin{equation}
\left[ v({\bf y}) v({\bf y}^{\prime}) \right] = \bar{f}
h^{-(d-1)} \delta_{{\bf y}, {\bf y}^{\prime}} .
\end{equation}
While deviations from the Gaussian distribution turn out to be
irrelevant for the asymptotic scaling behaviour long range
correlations of the velocity field~\cite{ALR} or a non-vanishing
mean shear~\cite{HB} change the universality class. The motion
of the particles in the velocity field is described be the
Liouvillean
\begin{eqnarray}
\fl \hat{L}_{\rm flow} = \frac{1}{h} \sum_{i} |v({\bf y}_{i})|
\left[ \hat{a}_{\rm A}^{\star}({\bf r}_{i} + {\rm sgn}(v({\bf
y}_{i})) h {\bf e}_{x}) - \hat{a}_{\rm A}^{\star}({\bf r}_{i})
\right] \hat{a}_{\rm A}({\bf r}_{i}) \nonumber \\
+ \frac{1}{h} \sum_{i} |v({\bf y}_{i})| \left[
\hat{a}_{\rm B}^{\star}({\bf r}_{i} + {\rm sgn}(v({\bf y}_{i}))
h {\bf e}_{x}) - \hat{a}_{\rm B}^{\star}({\bf r}_{i}) \right]
\hat{a}_{\rm B}({\bf r}_{i})
\end{eqnarray}
where ${\rm sgn}(v) = \pm 1$ is the sign function and ${\bf 
e}_{x}$ denotes the unit vector in positive $x$-direction.
In order to perform the average over the realizations of $v({\bf
y})$ it will be convenient to write $\hat{L}_{\rm flow}$ as a
sum, $\hat{L}_{\rm flow} = \hat{L}^{(-)} + \hat{L}^{(+)}$, where
\begin{eqnarray}
\fl \hat{L}^{(-)} = \frac{1}{2 h} \sum_{i} v({\bf y}_{i})
\left[ \hat{a}_{\rm A}^{\star}({\bf r}_{i} + h {\bf e}_{x}) -
\hat{a}_{\rm A}^{\star}({\bf r}_{i} - h {\bf e}_{x}) \right]
\hat{a}_{\rm A}({\bf r}_{i}) \nonumber \\
+ \frac{1}{2 h} \sum_{i} v({\bf y}_{i})
\left[ \hat{a}_{\rm B}^{\star}({\bf r}_{i} + h {\bf e}_{x}) -
\hat{a}_{\rm B}^{\star}({\bf r}_{i} - h {\bf e}_{x}) \right]
\hat{a}_{\rm B}({\bf r}_{i}) \label{L-}
\end{eqnarray}
is odd with respect to the velocity field and
\begin{eqnarray}
\fl \hat{L}^{(+)} = \frac{1}{2 h} \sum_{i} |v({\bf y}_{i})|
\left[ \hat{a}_{\rm A}^{\star}({\bf r}_{i} + h {\bf e}_{x}) +
\hat{a}_{\rm A}^{\star}({\bf r}_{i} - h {\bf e}_{x}) - 2
\hat{a}_{\rm A}^{\star}({\bf r}_{i})\right] \hat{a}_{\rm
A}({\bf r}_{i}) \nonumber \\
+ \frac{1}{2 h} \sum_{i} |v({\bf y}_{i})| \left[ \hat{a}_{\rm
B}^{\star}({\bf r}_{i} + h {\bf e}_{x}) + \hat{a}_{\rm
B}^{\star}({\bf r}_{i} - h {\bf e}_{x}) - 2 \hat{a}_{\rm
B}^{\star}({\bf r}_{i})\right] \hat{a}_{\rm B}({\bf r}_{i})
\label{L+}
\end{eqnarray}
depends only on the modulus $|v({\bf y})|$. The even part
$\hat{L}^{(+)}$ can be interpreted as a contribution to the
diffusion in $x$-direction with a ${\bf y}$-dependent diffusion
constant.

The formal solution of equation~(\ref{Liou}) is given by
\begin{equation}
|\Phi(t)\rangle = \exp(\hat{L} t) |\Phi(0)\rangle
\end{equation}
with $\hat{L} = \hat{L}_{\rm diff} + \hat{L}_{\rm reac} +
\hat{L}_{\rm flow}$.
In order to derive a functional integral representation for
the dynamics one uses the Trotter formula
\begin{equation}
\exp(\hat{L} t) = \lim_{n \rightarrow \infty} (1+\hat{L}
t/n)^{n}
\end{equation}
to rewrite the time evolution operator as a product of n
factors linear in $\hat{L}$. Inserting the identity operator
in a coherent state representation between the individual
factors~\cite{Pel,BLee,LC1} one obtains a path integral with
action
\begin{eqnarray}
\fl S[\tilde{a}, a; \tilde{b}, b] = h^{d} \int \d t \sum_{i} [
\tilde{a}({\bf r}_{i}, t) \partial_t a({\bf r}_{i}, t) +
\tilde{b}({\bf r}_{i}, t) \partial_t b({\bf r}_{i}, t) ]
\nonumber \\
- \int \d t L[\tilde{a}(t), a(t); \tilde{b}(t), b(t)] -
h^{d} \sum_{i} n_{0} \left( \tilde{a}_{\rm A}({\bf r}_{i}, 0) +
\tilde{a}_{\rm B}({\bf r}_{i}, 0) \right) \label{12}
\end{eqnarray}
where $\tilde{a}$, $a$, $\tilde{b}$ and $b$ are c-number
functions and $L[\tilde{a}(t), a(t); \tilde{b}(t), b(t)]$
can be obtained from $\hat{L}$ via the replacements
\begin{equation}
\eqalign{
\hat{a}_{\rm A}({\bf r}_{i}) \rightarrow h^{d} a({\bf r}_{i}, t)
\qquad & \hat{a}_{\rm A}^{\star}({\bf r}_{i}) \rightarrow 1 +
\tilde{a}({\bf r}_{i}, t) \\
\hat{a}_{\rm B}({\bf r}_{i}) \rightarrow h^{d} b({\bf r}_{i}, t)
& \hat{a}_{\rm B}^{\star}({\bf r}_{i}) \rightarrow 1 +
\tilde{b}({\bf r}_{i}, t) .
}
\end{equation}
In equation~(\ref{12}) $n_{0}$ denotes the initial density of
each species.

In order to calculate the average of correlation and response
functions with respect to disorder we use the effective action
$\bar{S}$ defined by
\begin{equation}
\exp(-\bar{S}[\tilde{a}, a; \tilde{b}, b]) = \left[
\exp(-S[\tilde{a}, a; \tilde{b}, b]) \right] .
\end{equation}
For each value of the coordinate ${\bf y}$ one has to calculate
an integral of the form
\begin{equation}
\int_{-\infty}^{\infty} d v({\bf y}) (2 \pi
\bar{f}/h^{d-1})^{-1/2} \exp\left( \frac{v({\bf y})^{2}}{2
\bar{f}/h^{d-1}} \right) \cdot \exp(- A v({\bf y}) - B |v({\bf
y})|) \label{av}
\end{equation}
where $A$ and $B$ are shorthand notations for the contributions
to $S$ coming from $\hat{L}^{(-)}$ and $\hat{L}^{(+)}$,
respectively. The integral~(\ref{av}) can be expressed in 
terms of error functions. We only need the approximation
\begin{equation}
(\protect\ref{av}) = \exp\left( \frac{1}{2} (\bar{f}/h^{d-1})
A^{2} - \sqrt{\frac{2}{\pi}} (\bar{f}/h^{d-1})^{1/2} B + \ldots
\right)
\end{equation}
since higher orders in $A$ and $B$ turn out to be irrelavant
for the asymptotic scaling behaviour. The first term in the
exponential function (proportional to $A^{2}$) gives rise to a
new interaction which is non-local with respect to time while
the second contribution (linear in $B$) modifies the diffusion
constant in the direction parallel to the flow. Neglecting all
irrelevant terms one arrives at
\begin{eqnarray}
\fl \bar{S}[\tilde{a},a; \tilde{b}, b] = \int \d t \int \d^{d}r
\left[ \tilde{a} (\partial_{t} a - D \triangle_{\bot} a - D_{\|}
\partial_{\|}^{2} a) \right. \nonumber \\
+ \left. \tilde{b} (\partial_{t} b - D \triangle_{\bot} b - D_{\|}
\partial_{\|}^{2} b) + k a b (\tilde{a}+\tilde{b}+\tilde{a} \tilde{b})
- n_{0} (\tilde{a} + \tilde{b}) \delta(t) \right] \nonumber \\
- \frac{\bar{f}}{2} \int \d^{d-1}y \left( \int \d t \int \d x
(\tilde{a} \partial_{\|}a + \tilde{b} \partial_{\|}b) \right)^{2} .
\label{actab}
\end{eqnarray}
Here the sum over lattice sites has been replaced by an integral
over the continuous variable ${\bf r} = (r_{\|}, {\bf r}_{\bot})
= (x,  {\bf y})$. This continuum model is appropriate for the
study of the long-time and large-distance behaviour of the
system.

Since the quantity $\psi = (a-b)/\sqrt{2}$ is closely related to
the conserved density difference of A-particles and B-particles
it is convenient to introduce the fields~\cite{LC1}
\begin{equation}
\fl \psi = \frac{1}{\sqrt{2}} (a-b) \qquad \tilde{\psi} =
\frac{1}{\sqrt{2}} (\tilde{a} - \tilde{b}) \qquad \phi =
\frac{1}{\sqrt{2}} (a+b) \qquad \tilde{\phi} = \frac{1}{\sqrt{2}}
(\tilde{a} + \tilde{b}) . \label{psi}
\end{equation}
After rescaling of time by the diffusion constant $D$ the action
becomes in terms of the new variables
\begin{eqnarray}
\fl \bar{S}[\tilde{\psi},\psi; \tilde{\phi}, \phi] = \int \d t \int
\d^{d}r \left[ \tilde{\psi} (\partial_{t} \psi - \triangle_{\bot}
\psi - \kappa \partial_{\|}^{2} \psi) + \tilde{\phi} (\partial_{t}
\phi - \triangle_{\bot} \phi - \kappa \partial_{\|}^{2} \phi)
\right. \nonumber \\
\left. + \lambda_{1} \tilde{\phi} (\phi^{2} - \psi^{2}) +
\lambda_{2} (\tilde{\phi}^{2} - \tilde{\psi}^{2}) (\phi^{2} -
\psi^{2}) - \sqrt{2} n_{0} \tilde{\phi} \delta(t) \right]
\nonumber \\
- \frac{f}{2} \int \d^{d-1}y \left( \int \d t \int \d x
(\tilde{\psi} \partial_{\|}\psi + \tilde{\phi} \partial_{\|}\phi)
\right)^{2} \label{act}
\end{eqnarray}
with the coupling coefficients $\lambda_{1} = k/(\sqrt{2} D)$,
$\lambda_{2} = k/(4 D)$, $f = \bar{f}/D^{2}$ and $\kappa =
D_{\|}/D$.

The action $\bar{S}$ includes two different types of
fluctuations.  The interaction proportional to 
$\lambda_{2}$ leads to anticorrelations neglected in the mean
field rate equations. In the case of the single species
annihilation $A+A \rightarrow \emptyset$ these anticorrelations
are responsible for the slow density decay $n(t) \sim t^{-d/2}$
in $d<2$~\cite{BLee}. The interaction proportional to
$f$ describes the effect of the quenched random velocity field
and gives rise to the superdiffusive motion of particles found in 
references \cite{BGKPR,JH}. A straightforward dimensional
analysis shows that $\lambda_{2}$ is relevant below
$d_{\lambda}=2$ while the upper critical dimension of the
disorder is $d_{f}=3$. As already pointed out in~\cite{LC1} this
does not mean that $\lambda_{2}$ may be neglected from the
outset if $d>d_{\lambda}$. For non-vanishing $\lambda_{2}$
coarse graining of the action $\bar{S}$ generates new
interactions which are relevant below four dimensions. Since
these interactions are located at the `time surface' $t=0$ they
modify the initial distribution of the fields $\psi$ and $\phi$.
An effective action appropriate for the analysis of fluctuation
effects in dimensions $d>d_{\lambda}$ includes all relevant
interactions which may be generated by the renormalization
group for $\lambda_{2}>0$.

We will use renormalization group improved perturbation theory
to derive results from the effective action for non-zero $f$.
Since this calculation is based on the results of
reference~\cite{JH} for non-interacting particles (i.e.
$\lambda_{1} = \lambda_{2} = 0$) it is useful to review the
renormalization group approach to this problem briefly in the
next section.

\section{Non-interacting particles in a random velocity field}
\label{nonint}

For $\lambda_{1} = \lambda_{2} = n_{0} = 0$ only Green functions
of the form
\begin{equation}
G_{M,N}(\{{\bf r}, t\}) =
\left.\left\langle \prod_{i=1}^{M} \psi({\bf r}_{i}, t_{i})
\tilde{\psi}(\tilde{{\bf r}}_{i}, \tilde{t}_{i}) \cdot
\prod_{j=M+1}^{M+N} \phi({\bf r}_{j}, t_{j})
\tilde{\phi}(\tilde{{\bf r}}_{j}, \tilde{t}_{j}) \right\rangle
\right|_{\lambda=0}
\label{GMN}
\end{equation}
with an equal number of insertions of $\psi$ ($\phi$) and
$\tilde{\psi}$ ($\tilde{\phi}$) are non-zero.
(Throughout this section the angular brackets indicate an
average with weight $\exp(-\bar{S})$ for $\lambda_{1} =
\lambda_{2} = n_{0} = 0$.)

Since $f$ is a relevant parameter below three dimensions the
asymptotic scaling behaviour of the Green functions cannot be
obtained from a na\"{\ii}ve perturbation expansion at finite
order in $f$. We therefore use the renormalization group to
improve the perturbation theory by a partial summation of the
perturbation series. Due to ultraviolet-divergencies at the upper
critical dimension $d_{f}=3$ the individual contributions to
the perturbation series have poles in $\epsilon = 3-d$.
In the minimal renormalization scheme these poles are absorbed
into renormalizations of coupling coefficients and fields.

\begin{figure}[b]
\caption{The only superficially divergent Feynman diagram of the
field theory defined by the action
$\bar{S}$ [equation~(\protect\ref{act})] with $\lambda_{1} =
\lambda_{2} = n_{0} = 0$. The full line with an arrow represents
the propagator~(\protect\ref{prop}) and the wavy line corresponds
to the correlator $f \delta({\bf r}_{\bot} - {\bf
r}_{\bot}^{\prime})$ of the velocity field.}
\label{fig1}
\vspace*{5mm}
\epsfxsize=160pt
\hspace*{50mm}\epsfbox{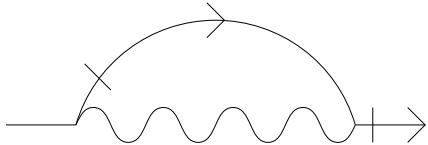}
\end{figure}

The propagator of the free (Gaussian) field theory is given
by
\begin{equation}
\langle \psi({\bf r}, t) \tilde{\psi}({\bf 0}, 0) \rangle_{0}
= \langle \phi({\bf r}, t) \tilde{\phi}({\bf 0}, 0) \rangle_{0}
= \int_{{\bf q}} \e^{\i {\bf q} {\bf r}} G({\bf q}, t)
\end{equation}
with the Fourier transform
\begin{equation}
G({\bf q}, t) = \Theta(t) \exp\left(-(q_{\bot}^{2} + \kappa 
q_{\|}^{2}) t\right) . \label{prop}
\end{equation}
Here $\Theta(t)$ is the Heaviside step function and the indices
`$\|$' and `$\bot$' indicate the respective directions parallel
and perpendicular to the flow.
Due to the strong anisotropy of the velocity field the theory
is superrenormalizable in the upper critical dimension~\cite{JH},
i.e. the number of superficially divergent diagrams is finite. 
In fact, it is sufficient to subtract the $\epsilon$-pole of the
diagram shown in figure~\ref{fig1} to render the field theory
finite at every order of the perturbation theory.
The required renormalization factor can be obtained from the
response function $G_{1,0}=G_{0,1}$ at one-loop order. A
straightforward calculation yields
\begin{eqnarray}
\fl \int \d^{d}r \exp(-\i q_{\|} r_{\|}) G_{1,0}({\bf r}, t) =
\exp(-\kappa q_{\|}^{2} t) \left[ 1 - \frac{2
f}{\epsilon (1+\epsilon/2) (4 \pi)^{1-\epsilon/2}} q_{\|}^{2}
t^{1+\epsilon/2} + \Or(f^{2} q_{\|}^{4}) \right]. \nonumber \\
\label{1loop}
\end{eqnarray}
To subtract the $\epsilon$-pole in this function we
introduce the renormalized diffusion constant $\kappa_{R} =
Z^{-1} \kappa$, where
\begin{equation}
Z = 1 - \frac{2 u}{\epsilon} \qquad
{\rm and} \qquad A_{\epsilon} f = \kappa_{R} u \mu^{\epsilon} .
\end{equation}
Here $u$ is a renormalized coupling coefficient, $A_{\epsilon} =
1/((4 \pi)^{1-\epsilon/2} (1+\epsilon/2))$ is a geometrical
factor and $\mu$ denotes an external momentum scale.

Since the bare Green function is independent of $\mu$ its
derivative at fixed bare parameters vanishes, i.e.
\begin{equation}
\mu \left.\frac{\d}{\d \mu}\right|_{0} G_{M,N}(\{{\bf r}, t\};
\kappa, f) = 0 .
\end{equation}
Expressing $\kappa$ and $f$ in terms of $\kappa_{R}$, $u$ and
$\mu$ we obtain for the renormalized Green function the
renormalization group equation (RGE)
\begin{equation}
\left[ \mu \partial_{\mu} - \zeta(u) \kappa_{\rm R}
\partial_{\kappa_{\rm R}} + \beta(u) \partial_{u} \right]
G_{M,N}(\{{\bf r}, t\}; \kappa_{R}, u; \mu) = 0 . \label{RGE}
\end{equation}
Due to superrenormalizability the Wilson functions
\begin{equation}
\zeta(u) = \mu \left.\frac{\d}{\d \mu}\right|_{0} \ln Z = 2 u
\qquad \beta(u) = \mu \left.\frac{\d}{\d \mu}\right|_{0} u =
u (-\epsilon + 2 u)
\end{equation}
are exact at every order of the perturbation theory.
Solving the RGE by characteristics one finds
\begin{equation}
G_{M,N}(\{{\bf r}, t\}; \kappa_{R},
u; \mu) = G_{M,N}(\{{\bf r}, t\};
Y(l) \kappa_{R}, \bar{u}(l); \mu l) \label{flow}
\end{equation}
where
\begin{equation}
\fl l \frac{\d}{\d l} \ln Y(l) = -\zeta(\bar{u}(l))
\qquad l \frac{\d}{\d l} \bar{u}(l) = \beta(\bar{u}(l))
\qquad Y(1) = 1 \qquad \bar{u}(1) = u. \label{29}
\end{equation}
For $l \rightarrow 0$ the scale dependent coupling constant
$\bar{u}(l)$ tends to the fixed point $u_{\star} = \epsilon/2$
and $Y(l) \simeq Y_{\star} l^{-\epsilon}$ (with a non-universal
scaling factor $Y_{\star}$). The asymptotic scaling behaviour
of the Green functions follows from equation~(\ref{flow}) and
dimensional analysis. The result reads (for $l \rightarrow 0$)
\begin{eqnarray}
\fl G_{M,N}(\{{\bf r}, t\}; \kappa_{R}, u; \mu) \simeq
\left(\mu^{d}  (Y_{\star} \kappa_{R})^{-1/2} l^{d+\epsilon/2}
\right)^{M+N} \nonumber \\
\times  G_{M, N}\left(\left\{\mu (Y_{\star} \kappa_{R})^{-1/2}
l^{1+\epsilon/2} r_{\|}, \mu l {\bf r}_{\bot}, \mu^{2} l^{2}
t\right\}; 1, u_{\star}; 1\right) . \label{green}
\end{eqnarray}
This equation shows that anomalous scaling dimensions of
the variables and fields are given by
\begin{equation}
\fl r_{\bot} \sim l^{-1} \qquad r_{\|} \sim l^{-(1+\epsilon/2)}
\qquad t \sim l^{-2} \qquad \tilde{\psi} \psi \sim l^{d +
\epsilon/2} \qquad \tilde{\phi} \phi \sim l^{d + \epsilon/2} .
\label{scal}
\end{equation}

The asymptotic form of the Green functions~(\ref{green}) depends
on the arbitrary momentum scale $\mu$ and the non-universal
amplitude $Y_{\star}$. A combination of these parameters
occuring in~(\ref{green}) can be expressed by a quantity
that is accessible to experiments or simulations~\cite{BGKPR}.
Consider the random walk of a particle starting at time $t=0$
at the point ${\bf r} = {\bf 0}$. The mean square
displacement of the particle in $r_{\|}$-direction averaged over
the realizations of the velocity field reads
\begin{equation}
X(t)^{2} = \int \d^{d}r G_{1,0}({\bf r}, t) r_{\|}^{2} .
\end{equation}
Using the one-loop result~(\ref{1loop}) and
equation~(\ref{green}) with $l=(\mu^{2} t)^{-1/2}$ it is easy to
show that (for $t \rightarrow \infty$)
\begin{equation}
X(t) \simeq \sigma t^{1/2 + \epsilon/4} \qquad {\rm with} \qquad
\sigma = \sqrt{2} (\kappa_{R} Y_{\star})^{1/2} \mu^{\epsilon/2}
\left[1 + \Or(\epsilon^{2}) \right] .
\end{equation}
In the following sections we will always use the parameter
$\sigma$ instead of $Y_{\star}$. Due to the strong anisotropy of
the velocity field it is also possible to calculate $X(t)$
directly from equation~(\ref{1loop}) without application of the
renormalization group. Since the $n^{\rm th}$ order contribution
to $G_{1, 0}$ is proportional to $q_{\|}^{2n}$ and $X(t)^{2}$ is
proportional to the derivative of $G_{1, 0}$ with respect to
$q_{\|}^{2}$ at ${\bf q} = {\bf 0}$ the one-loop result is
sufficient to obtain the function $X(t)$ exactly:
\begin{equation}
X(t)^{2} = 2 \kappa t + \frac{8 f}{\epsilon (2+\epsilon)
(4 \pi)^{1-\epsilon/2}} t^{1 + \epsilon/2} . \label{34}
\end{equation}
For $\epsilon = 1$ this result was first given by Bouchaud
\etal~\cite{BGKPR}.

For $d = 3$ ($\epsilon = 0$) an ultraviolet cut-off $\Lambda$
is required to regularize the singularity in equation~(\ref{34}).
In this case the mean sqare displacement is asymptotically
given by
\begin{equation}
X(t)^{2} \simeq \frac{f}{2 \pi} t \ln(\Lambda^{2} t) .
\label{d=3}
\end{equation}

\section{Effective action below three dimensions} \label{Seff}

The effect of the $\lambda_{2}$-vertex in equation~(\ref{act})
on the long-time behaviour of a reaction diffusion system
without velocity field can be described by interactions located
at the `time surface' $t=0$. In reference~\cite{LC1} it was
shown by dimensional analysis that an interaction of the type
\begin{equation}
\int \d^{d}r \frac{1}{m! n!} \Delta_{m,n} \left.\tilde{\psi}^{m}
\tilde{\phi}^{n}\right|_{t=0}
\label{ini}
\end{equation}
is relevant below $d_{m,n} = 2(m+n)/(m+n-1)$ dimensions.
($\Delta_{1,0}$ and $\Delta_{0,1}$ are relevant in any
dimension.) The critical dimension of $\Delta_{m,n}$ is changed
if the motion of the particles is not purely diffusive, e.g. if
the particles are subject to a linear shear flow~\cite{HB}.
In order to see how the values of $d_{m,n}$ are changed by the
random velocity field consider the expansion of $\langle \phi(t)
\rangle$ in powers of $\lambda_{1}$, $\lambda_{2}$ and $f$.
For $d < d_{f} = 3$ ($d > d_{f}$) the contributions to this
expansion at a given order in $\lambda_{1}$ and $\lambda_{2}$
become more important (less important) for large $t$ if we
increase the order in $f$. Therefore the long-time behaviour of
$\langle \phi(t) \rangle$ for $d > 3$ is dominated by the
diffusive motion of the particles and we may set $f=0$. Below 
three dimensions we have to retain all orders in $f$ to determine
the asymptotic decay of the density. In this case
the many-particle response functions~(\ref{GMN}) play the same
role as the Gaussian propagotor~(\ref{prop}) in the case $f=0$.
It is clear that it is not possible to sum the the power series
in $f$ exactly but the renormalization group allows us a partial
summation and gives the correct scaling behaviour.

We first determine the scaling dimension of $\lambda_{2}$ for a 
non-vanishing velocity field below three dimensions. Using the
anomalous scaling dimensions given in~(\ref{scal}) one finds
$\lambda_{2} \sim l^{2-d-\epsilon/2} = l^{(1-d)/2}$, i.e.
$\lambda_{2}$ is irrelevant for $d > 1$. In the same way we
obtain
\begin{equation}
\lambda_{1} \psi \sim l^{2} \qquad \lambda_{1} \phi \sim l^{2}
\qquad \lambda_{1}^{-1} \tilde{\psi} \sim l^{d-2+\epsilon/2}
\qquad \lambda_{1}^{-1} \tilde{\phi} \sim l^{d-2+\epsilon/2}
\label{scall}
\end{equation}
and the dimension of the initial term
\begin{equation}
\lambda_{1}^{m+n} \Delta_{m,n} \sim l^{(1-m-n)(d+\epsilon/2)
+ 2 (m+n)} .
\end{equation}
In $d=3$ the vertices with $m+n < 3$ are relevant. For $m+n \geq
3$ the initial coupling $\Delta_{m,n}$ is relevant below
$d_{m,n} = (m+n+3)/(m+n-1)$ dimensions. However, we will show
below that the only initial couplings which are generated by the
coarse graining procedure employed in reference~\cite{LC1} are
$\Delta_{2, 0}$ and $\Delta_{0, 2}$ (in addition to $\Delta_{0,
1} = \sqrt{2} n_{0}$), and these couplings are given by
$\Delta_{2, 0} = -\Delta_{0, 2} = n_{0}$ at every order in
$\lambda_{2}$.

\begin{figure}[b]
\caption{Contributions to the effective coupling coefficient
$\lambda_{2, {\rm eff}}$ to third order in $\lambda_{2}$.
The hatched rectangles represent the Green functions $G_{M,N}$
[equation~(\protect\ref{GMN})], where $M$ ($N$) is the number
of broken lines (full lines) running into the rectangle.
The direction of each line is indicated by an arrow and
reflects causality, i.e. each line points into the direction
of larger time arguments.}
\epsfxsize=360pt
\vspace*{5mm}\hspace*{20mm}\epsfbox{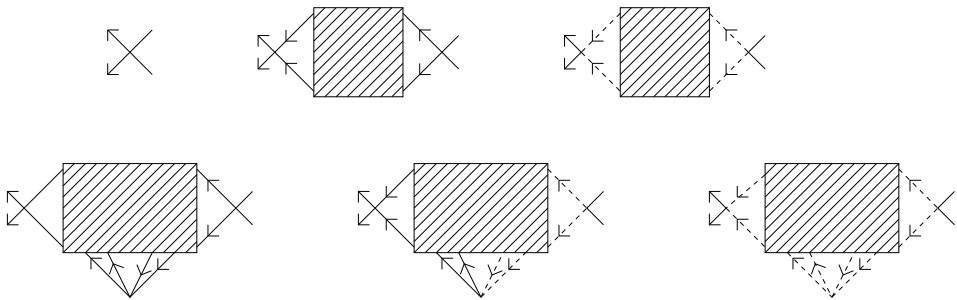}
\label{fig2}
\end{figure}

Since the coupling coefficient $\lambda_{2}$ is irrelevant
for $d>1$ one has to introduce a cut-off wave number $\Lambda$ to
avoid ultraviolet-divergencies in the perturbation series. The
cut-off is required to regularize the diagrams shown in
figure~\ref{fig2} which contribute to an effective coupling
constant $\lambda_{2,{\rm eff}}$. The Green functions $G_{M,N}$
considered in the previous section occur in these diagrams as
subintegrals (represented by rectangles). As in the previous
section we calculate these subintegrals without cut-off since
they are ultraviolet-convergent for $d<3$. After summation of
the subdiagrams (using the renormalization group) the cut-off
is introduced to perform the remaining integrations. If the
cut-off wave number is not too large [an estimate based on
equation~(\ref{34}) with $t \sim \Lambda^{-2}$ gives
$\Lambda^{\epsilon} \ll f/(\kappa \epsilon)$] the scaling
behaviour~(\ref{green}) still holds on the momentum scale of
$\Lambda$. In this case the effective coupling has the scaling
form
$\lambda_{2,{\rm eff}} = \lambda_{2} F_{\lambda}(\lambda_{2}
\Lambda^{(d-1)/2}/\sigma)$ .
In the same way coarse graining leads to an effective coupling
coefficient $\lambda_{1, {\rm eff}} = \lambda_{1}
F_{\lambda}(\lambda_{2} \Lambda^{(d-1)/2}/\sigma)$ (with the same
scaling function $F_{\lambda}$).
Evaluating the diagrams depicted in figure~(\ref{fig2}) by 
renormalization group improved pertubation theory it is possible
to compute the coefficients of the Taylor expansion of
$F_{\lambda}$ in an $\epsilon$-expansion. [Due to the scaling
behaviour~(\ref{green}) of the response functions the diagrams
are (infrared-) finite for $d > 1$ in the limit of zero external
momentum and frequency.] However, this is not very interesting
since the result strongly depends on the type of the cut-off.

The initial coupling $\Delta_{2, 0}$ is more interesting since
it is known to determine (at least for systems without random
velocity field) the amplitude of $\langle \phi(t)
\rangle$~\cite{LC1}. For the calculation of $\Delta_{2, 0}$ it is
convenient to shift the field $\phi$ by the mean field density
\begin{equation}
\Phi_{\rm mf}(t) = \Phi_{0} \left( 1 + \lambda_{1} \Phi_{0} t
\right)^{-1} \label{phimf}
\end{equation}
with the initial value $\Phi_{0} = \sqrt{2} n_{0}$. The shifted
action $\bar{S}[\tilde{\psi}, \psi; \tilde{\phi}, \Phi(t) + \phi]$
consists of a Gaussian part
\begin{eqnarray}
\fl \bar{S}_{G}[\tilde{\psi}, \psi; \tilde{\phi}, \phi; \Phi(t)]
= \int \d t \int \d^{d}r \left[ \tilde{\psi} (\partial_{t}
\psi - \triangle_{\bot} \psi - \kappa \partial_{\|}^{2} \psi)
\right. \\
\left. + \tilde{\phi} \left(\partial_{t} \phi - \triangle_{\bot}
\phi - \kappa \partial_{\|}^{2} \phi + 2\lambda_{1} \Phi_{\rm
mf}(t) \phi\right)\right] \nonumber
\end{eqnarray}
and higher order terms summarized in
\begin{eqnarray}
\fl \bar{S}_{\rm int}[\tilde{\psi}, \psi; \tilde{\phi}, \phi;
\Phi(t)] = \int \d t \int \d^{d}r \left[ \lambda_{1} \tilde{\phi}
(\phi^{2}-\psi^{2}) + \lambda_{2} (2 \Phi_{\rm mf}(t) \phi +
\Phi_{\rm mf}(t)^{2})
(\tilde{\phi}^{2} - \tilde{\psi}^{2}) \right. \label{sint} \\
\left. + \lambda_{2} (\tilde{\phi}^{2} - \tilde{\psi}^{2}) (\phi^{2}
- \psi^{2}) \right] - \frac{f}{2} \int \d^{d-1}y \left( \int
\d t \int \d x (\tilde{\psi} \partial_{\|}\psi + \tilde{\phi}
\partial_{\|}\phi) \right)^{2} . \nonumber
\end{eqnarray}
Since $\bar{S}_{G}$ is independent of $f$ the Gaussian (tree)
propagators are the same as in reference~\cite{LC1}, i.e.
\begin{eqnarray}
\fl G_{\psi}({\bf q}; t, t^{\prime}) = \int \d^{d}r \e^{-\i {\bf
q} {\bf r}} \langle \psi({\bf r}, t) \tilde{\psi}({\bf 0},
t^{\prime}) \rangle_{\rm G} = G({\bf q}; t-t^{\prime})\\
\fl G_{\phi}({\bf q}; t, t^{\prime}) = \int \d^{d}r \e^{-\i {\bf
q} {\bf r}} \langle \phi({\bf r}, t) \tilde{\phi}({\bf 0},
t^{\prime}) \rangle_{\rm G} = \left( \frac{1 + \lambda_{1}
\Phi_{0} t^{\prime}}{1 + \lambda_{1} \Phi_{0} t} \right)^{2}
G({\bf q}; t-t^{\prime}) \label{gphi}
\end{eqnarray}
where $G({\bf q}; t)$ is the propagator defined in
equation~(\ref{prop}). Corrections to the mean field density
can now be calculated in a diagramatic expansion around the
Gaussian action $\bar{S}_{G}$ treating the interactions in
$\bar{S}_{\rm int}$ as perturbations.

\begin{figure}[b]
\caption{General form of the contributions to $\Delta_{2, 0}$.
The full and broken bold lines represent the 
propagators~$G_{\phi}$ and~$G_{\psi}$, respectively, and the
zigzag line corresponds to the mean field density $\Phi_{\rm
mf}(t)$. At lowest order in $\lambda_{2}$ the bubbles in the
graphs~(b) and~(c) contain only the graph~(a) as a subdiagram.
The vertex proportional to $\Phi_{\rm mf}(t)$ in the
diagrams~(c) and~(e) is generated by the shift of the field
$\phi$ [see equation~(\protect\ref{sint})].}
\epsfxsize=300pt
\vspace*{5mm}\hspace*{25mm}\epsfbox{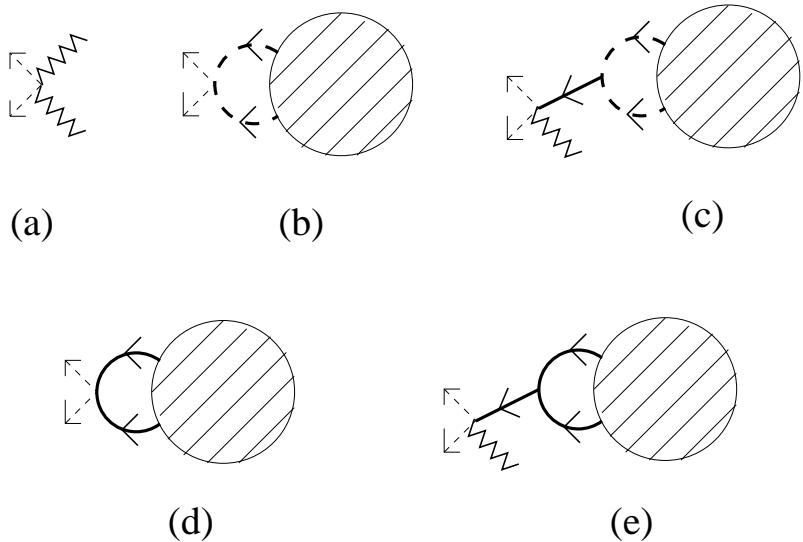}
\label{fig3}
\end{figure}
Following the lines taken in reference~\cite{LC1} we replace
the irrelevant parameter $\lambda_{2}$ by effective initial
couplings of the form~(\ref{ini}). The diagrams contributing to
$\Delta_{m,n}$ have $(m+n)$ external legs each of which is
associated with one of the vertices proportional to
$\lambda_{2}$. Figure~\ref{fig3} shows the general form of the
contributions to $\Delta_{2,0}$.
Since the function $\Phi_{\rm mf}(t)^{2}$ is damped on
time scales which are large compared to $(\lambda_{1}
\Phi_{0})^{-1}$ its effect on the long-time behaviour of the
field theory can be described by a $\delta$-function. We may
therefore write
\begin{equation}
\lambda_{2} \Phi_{\rm mf}(t)^{2} \tilde{\psi}^{2} \simeq
\lambda_{2} \left[ \int_{0}^{\infty} \d t^{\prime} \Phi_{\rm
mf}(t^{\prime})^{2} \right] \delta(t) \tilde{\psi}^{2} =
\frac{\lambda_{2}}{\lambda_{1}} \Phi_{0} \delta(t) \tilde{\psi}^{2}
\end{equation}
and obtain $\Delta_{2, 0} = 2 \lambda_{2} \Phi_{0}/ \lambda_{1}
+ \Or(\lambda_{2}^{2})$. In the same way higher order
contributions to $\Delta_{2, 0}$ can be calculated by integrating
the diagrams~(b--e) in figure~\ref{fig3} with respect to the
time argument associated with the leftmost vertex. However, the
diagrams~(b) and~(c) cancel, and the same is true for the
diagrams~(d) and~(e). In order to see this we write the 
contribution of the graph~\ref{fig3}(b) in the form
\begin{equation}
{\rm graph~\protect\ref{fig3} (b)} =  \int_{0}^{\infty}
\d t (-\lambda_{2}) f(t)
\end{equation}
where $t$ is the time argument carried by the leftmost vertex,
and the function $f(t)$ represents the subdiagrams in the
hatched bubble and the loop integration. The same function $f(t)$
occurs in the contribution~\ref{fig3}(c) which is given by
\begin{eqnarray}
{\rm graph~\protect\ref{fig3} (c)} &= \int_{0}^{\infty} \d t
2\lambda_{2} \Phi_{\rm mf}(t) \int_{0}^{t} \d t^{\prime}
G_{\phi}({\bf q} = {\bf 0}; t, t^{\prime}) \lambda_{1}
f(t^{\prime}) \nonumber \\
&= 2 \lambda_{1} \lambda_{2} \Phi_{0}  \int_{0}^{\infty} \d
t^{\prime} \int_{t^{\prime}}^{\infty} \d t  \frac{(1+\lambda_{1}
\Phi_{0} t^{\prime})^{2}}{(1 + \lambda_{1} \Phi_{0} t)^{3}} 
f(t^{\prime}) \\
&= \int_{0}^{\infty} \d t^{\prime} \lambda_{2} f(t^{\prime}) .
\nonumber
\end{eqnarray}
Therefore the contributions~(b) and~(c) cancel. Analogously
one can show that the sum of the diagrams~(d) and~(e)
vanishes. Since the same line of arguments applies if the
hatched bubbles in figure~\ref{fig3} have additional external 
legs\footnote{Due to causality each diagram which contributes
to the initial couplings contains at least one pair of
$\tilde{\psi}$- or $\tilde{\phi}$- legs coming from a single
$\lambda_{2}$ vertex. This is the vertex explicitely shown in
figure~\protect\ref{fig3}.}, all effective interactions
$\Delta_{m, n}$ with $m+n > 2$ are zero, and the only
non-vanishing initial couplings are given by $\Delta_{2, 0} =
-\Delta_{0, 2} = 2 \lambda_{2} \Phi_{0}/ \lambda_{1} = n_{0}$.
In the appendix of reference~\cite{LC1} a non-zero correction to
$\Delta_{2, 0}$ of the order $\Or(\lambda_{2}^{2} n_{0}^{d/2})$
was obtained because the diagrams of the form~(c) and~(e) were
incorrectly thought to be accounted for by taking $\lambda_{1}$
to $\lambda_{\rm eff}$.

The cancellation of diagrams suggests that there should be a
simpler way to calculate the initial coupling coefficients. In
fact, the effective action can be derived without using Feynman
diagrams. This formal derivation exploits the equation of
motion
\begin{equation}
\int D[\tilde{\psi}, \psi; \tilde{\phi}, \phi]
\frac{\delta}{\delta \tilde{\psi}({\bf r}, t)}
\exp(-S[\tilde{\psi}, \psi; \tilde{\phi}, \phi]) = 0
\end{equation}
and the corresponding equation with a functional derivative
with respect to $\tilde{\phi}$. The explicit form of the
equations of motion (which are valid after insertion into
averages) is given by
\begin{equation}
\fl \partial_{t} \psi - \triangle_{\bot} \psi -\kappa
\partial_{\|}^{2} \psi - 2 \lambda_{2} \tilde{\psi} (\phi^{2} -
\psi^{2})
- f (\partial_{\|} \psi) \int \d t^{\prime} \int \d x^{\prime}
\left( \tilde{\psi} \partial_{\|} \psi + \tilde{\phi}
\partial_{\|} \phi \right)
= 0 \label{eqmo1}
\end{equation}
and
\begin{eqnarray}
\fl \partial_{t} \phi - \triangle_{\bot} \phi -\kappa
\partial_{\|}^{2} \phi + \lambda_{1} (\phi^{2} - \psi^{2}) + 2
\lambda_{2} \tilde{\phi} (\phi^{2} - \psi^{2}) -
\Phi_{0} \delta(t) \nonumber \\
- f (\partial_{\|} \phi) \int \d t^{\prime} \int \d x^{\prime}
\left( \tilde{\psi} \partial_{\|} \psi + \tilde{\phi}
\partial_{\|} \phi \right)
= 0 . \label{eqmo2}
\end{eqnarray}
Averaging~(\ref{eqmo2}) one finds
\begin{equation}
\partial_{t} \Phi(t) + \lambda_{1} \langle \phi({\bf r}, t)^{2}
- \psi({\bf r}, t)^{2} \rangle = \Phi_{0} \delta(t) \label{56}
\end{equation}
where $\Phi(t) = \langle \phi({\bf r}, t) \rangle$ is the exact
density. (The equal time averages $\langle \tilde{\phi} \psi^{2}
\rangle$ and $\langle \tilde{\phi} \phi^{2} \rangle$ are zero due
to the prepoint time discretization used in the derivation of the
action.) Integration of equation~(\ref{56}) over t yields
\begin{equation}
\lambda_{1} \int_{0}^{\infty} \d t \langle \phi({\bf r}, t)^{2}
- \psi({\bf r}, t)^{2} \rangle = \Phi_{0} .
\end{equation}
The initial fluctuations generated by the irrelevant
$\lambda_{2}$ coupling can now be obtained by the replacement
\begin{equation}
2 \lambda_{2} (\phi^{2} - \psi^{2}) \longrightarrow
2 \lambda_{2} \delta(t)  \int_{0}^{\infty} \d t^{\prime}
\langle \phi({\bf r}, t^{\prime})^{2} - \psi({\bf r},
t^{\prime})^{2} \rangle = n_{0} \delta(t)
\end{equation}
in the equations of motion~(\ref{eqmo1}) and~(\ref{eqmo2}).
The effective equations of motion derived in this way are
equivalent to the effective action with the initial vertices
$\Delta_{m, n}$ calculated above.

\section{Asymptotic decay of the density} \label{decay}

In order to calculate the long-time behaviour of $\langle\phi(t)
\rangle$ we start from the effective action $\bar{S}_{\rm eff} =
\bar{S}_{\rm reac} + S_{\rm ini}$, where
\begin{eqnarray}
\fl \bar{S}_{\rm reac}[\tilde{\psi}, \psi; \tilde{\phi}, \phi] =
\int \d t \int \d^{d} r \left[\tilde{\psi} \left(\partial_{t} \psi
- \triangle_{\bot} \psi -\kappa \partial_{\|} \psi \right)
+ \tilde{\phi} \left(\partial_{t} \phi - \triangle_{\bot}
\phi -\kappa \partial_{\|} \phi \right) \right. \nonumber \\
\left. + \lambda_{1} \tilde{\phi}  (\phi^{2}-\psi^{2}) \right]
- \frac{f}{2} \int \d^{d-1}y \left( \int
\d t \int \d x (\tilde{\psi} \partial_{\|}\psi + \tilde{\phi}
\partial_{\|}\phi) \right)^{2}
\end{eqnarray}
and
\begin{equation}
S_{\rm ini}[\tilde{\psi}, \tilde{\phi}] = -\int \d^{d}r
\left[ \Phi_{0} \tilde{\phi} + \frac{1}{2} n_{0} \left(
\tilde{\psi}^{2} - \tilde{\phi}^{2} \right) \right]_{t=0} .
\end{equation}
At this point it is convenient to reintroduce the
Gaussian velocity field $v({\bf y})$ (with $[v({\bf y})]=0$ and
$[v({\bf y}) v({\bf y}^{\prime})] = f \delta({\bf y} - {\bf
y}^{\prime})$ as in section~\ref{model}) to replace
$\bar{S}_{\rm reac}$ in favour of a $v$-dependent action which is 
local with repect to time. This new action is equivalent to the
equations
\begin{eqnarray}
\partial_{t} \psi - \triangle_{\bot} \psi - \kappa
\partial_{\|}^{2} \psi + v({\bf y}) \partial_{\|} \psi = 0
\label{eqpsi} \\
\partial_{t} \phi - \triangle_{\bot} \phi - \kappa
\partial_{\|}^{2} \phi + v({\bf y}) \partial_{\|} \phi 
+ \lambda (\phi^{2} - \psi^{2}) = 0 \label{eqphi}
\end{eqnarray}
where $\psi$ and $\phi$ are now {\em real} fields.
The solutions $\psi$, $\phi$ have to be averaged with respect to
both the initial conditions and the realizations of the velocity
field. Since the initial state defined by $S_{\rm ini}$ and the
distribution of $v({\bf y})$ are homogeneous the averages of the
spatial derivatives in~(\ref{eqpsi},\ref{eqphi})
vanish.
Averaging equation~(\ref{eqphi}) we find
\begin{equation}
\fl \dot{\Phi}(t) + \lambda \left( \Phi(t)^{2} - \langle
\psi({\bf r}, t)^{2} \rangle + C(t) \right) = 0 \qquad {\rm
where} \qquad C(t) =  \left\langle (\phi({\bf r}, t)-\Phi(t))^{2}
\right\rangle . \label{Phi(t)}
\end{equation}
To obtain the asymptotic solution of this equation one
first has calculate $\langle \psi({\bf r}, t)^{2} \rangle$ for
large $t$. Later in this section we will also need higher moments
$\langle \psi^{2n} \rangle$ with $n= 1, 2\ldots$. Since in
equation~(\ref{eqpsi}) the dynamics of $\psi$ is decoupled from
the reaction process the moments of $\psi$ may be calculated with
the Green functions~(\ref{GMN}). In this way we get
\begin{eqnarray}
\fl \langle \psi({\bf r}, t)^{2n} \rangle = \frac{1}{2^{n} n!}
n_{0}^{n} \int \d^{d}r_{1}^{\prime} \ldots \int
\d^{d}r_{n}^{\prime} \left.\left\langle \psi({\bf r}, t)^{2 n}
\tilde{\psi}({\bf r}_{1}^{\prime}, 0)^{2} \ldots \tilde{\psi}({\bf
r}_{n}^{\prime}, 0)^{2} \right\rangle \right|_{\lambda=0}
\label{58} \\
\lo \simeq {\rm const} \times \left(\frac{n_{0}}{\sigma}
t^{-(d+3)/4} \right)^{n} \nonumber
\end{eqnarray}
where the scaling form~(\ref{green}) has been used.
[A simple power counting as in reference~\cite{JH} shows that
insertions of composite fields such as $\psi^{2n}$ in
equation~(\ref{58}) require no additional renormalizations.]

\begin{figure}[b]
\caption{Contributions to $\langle \psi^{2} \rangle$ at first
order in $\epsilon$. The full and empty circles represent the
initial vertex $\Delta_{2, 0}$ and the field $\psi^{2}$,
respectively. A solid line with an arrow corresponds to the
Gaussian propagator~(\protect\ref{prop}).}
\epsfxsize=300pt
\vspace*{5mm}\hspace*{25mm}\epsfbox{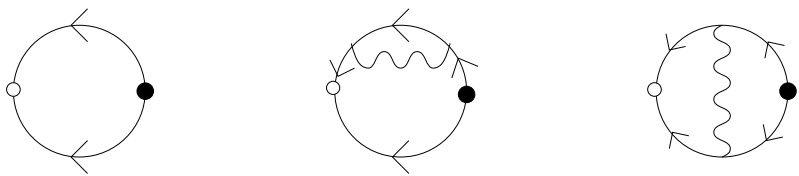}
\label{fig4}
\end{figure}

To obtain the amplitude of the second moment $\langle \psi^{2}
\rangle$ at first order in $\epsilon$ one has to calculate the
diagrams shown in figure~\ref{fig4}. The result reads in terms
of the unrenormalized coupling constants
\begin{equation}
\fl \langle \psi({\bf r}, t)^{2} \rangle = n_{0} \kappa^{-1/2}
(8 \pi t)^{-d/2} \left[ 1 - \frac{f t^{\epsilon/2}}{(4 \pi)^{1
-\epsilon/2} \kappa} \left(\frac{1}{\epsilon} - \frac{3}{2} \ln 2
+ \Or(\epsilon) \right) + \Or(f^{2}) \right] .
\end{equation}
Upon application of the renormalization group this becomes (for
large $t$)
\begin{equation}
\langle \psi({\bf r}, t)^{2} \rangle \simeq n_{0}
\frac{\sqrt{2}}{(8 \pi)^{d/2} \sigma} t^{-(d+3)/4} \left[
1 + \frac{\epsilon}{4} \left(3 \ln 2 - 1 \right) +
\Or(\epsilon^{2}) \right] .
\end{equation}
We are now in a position to determine the asymptotic decay of
$\Phi(t)$ in a similar way as in the case $v({\bf y}) \equiv
0$~\cite{LC1}:

We first solve equation~(\ref{Phi(t)}) for $C(t)=0$, i.e. we are
looking for a function $\Phi_{\rm u}(t)$ satisfying
\begin{equation}
\dot{\Phi}_{\rm u}(t) + \lambda \left( \Phi_{\rm u}(t)^{2} -
\langle \psi({\bf r}, t)^{2} \rangle \right) = 0.
\end{equation}
Since $\langle \psi^{2} \rangle = {\rm cst} \times t^{-(d+3)/4}$
for $t \rightarrow \infty$ (with $(d+3)/4 < 2$) the asymptotic
solution of this equation is given by $\Phi_{\rm u}(t) \simeq
\sqrt{{\rm cst}} \times t^{-(d+3)/8}$. Using the positivity of
$C(t)$ one can show~\cite{LC1} that $\Phi_{\rm u}(t)$ is an upper
bound for $\Phi(t)$.
To derive a lower bound $\Phi_{\rm l}(t) \leq \Phi(t)$
we use the fact that $a = (\phi + \psi)/\sqrt{2}$ and $b = (\phi
- \psi)/\sqrt{2}$ are real densities which satisfy the equations
of motion
\begin{equation}
\eqalign{
\partial_{t} a - \triangle_{\bot} a - \kappa \partial_{\|}^{2} a
+ v({\bf y}) \partial_{\|} a + \sqrt{2} \lambda a b =& 0 \\
\partial_{t} b - \triangle_{\bot} b - \kappa \partial_{\|}^{2} b
+ v({\bf y}) \partial_{\|} b + \sqrt{2} \lambda a b =& 0 .
}
\end{equation}
For non-negative initial values the densities remain non-negative
for all $t$. This implies $\phi^{2}-\psi^{2} = 2 a b \geq 0$
and leads to the lower bound $\Phi_{\rm l}(t) = \langle |\psi|
\rangle$.

\begin{figure}[b]
\caption{Non-zero contributions to $c_{4}(t)$ at second
order in $f$~(b-d). The first order diagram~(a) vanishes.}
\epsfxsize=350pt
\vspace*{5mm}\hspace*{17mm}\epsfbox{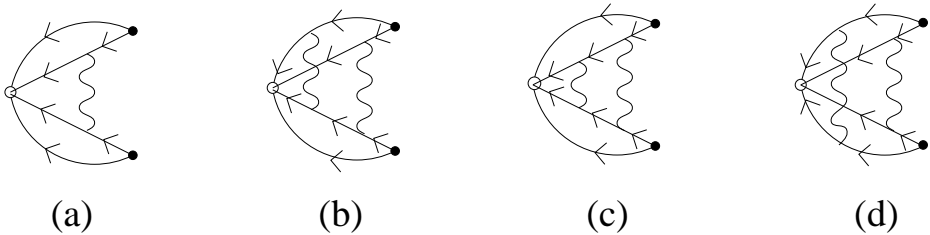}
\label{fig5}
\end{figure}

Although the initial distribution of $\psi$ is Gaussian, the
velocity field generates for $t>0$ higher cumulants in the
distribution ${\cal P}_{t}$ of $\psi({\bf r}, t)$. We perform a
cumulant expansion for this distribution to calculate $\langle
|\psi| \rangle$.
Since higher cumulants are generated by the disorder vertex $f$
(or its renormalized counterpart $v$ with fixed point value
$v_{\star} = \epsilon/2$) this amounts to an $\epsilon$-expansion
for ${\cal P}_{t}(\psi)$.
The Fourier transform $\tilde{{\cal P}}_{t}$ of the distribution
with respect to $\psi$ can be written as
\begin{equation}
\tilde{{\cal P}}_{t}(h) = \int_{-\infty}^{\infty} \d \psi
{\cal P}_{t}(\psi) \e^{\i h \psi} = \exp\left(
\sum_{n=1}^{\infty} \frac{c_{2n}(t)}{(2n)!} (-1)^{n} h^{2n}
\right) \label{four}
\end{equation}
where $c_{2n}(t)$ denotes the connected part (cumulant) of the
$(2n)^{\rm th}$ moment of $\psi$, e.g.
\begin{equation}
c_{2}(t) = \langle \psi({\bf r}, t)^{2} \rangle \qquad
c_{4}(t) = \langle \psi({\bf r}, t)^{4} \rangle - 3 c_{2}(t)^{2}
\qquad {\rm etc.}
\end{equation}
For a Gaussian distribution all $c_{m}(t)$ with $m > 2$ vanish.
Equation(\ref{four}) can be used to calculate $\langle |\psi|
\rangle$ in an expansion in $c_{2n}(t)$, $n=4, 6, \ldots$:
\begin{eqnarray}
\fl \langle |\psi({\bf r}, t)| \rangle = \int_{-\infty}^{\infty}
\d \psi |\psi| \int_{-\infty}^{\infty} \frac{\d h}{2 \pi}
\tilde{{\cal P}}_{t}(h) \e^{-\i h \psi} \nonumber \\
\lo = \sqrt{\frac{2 c_{2}(t)}{\pi}} \left[ 1 - \frac{1}{4!}
\frac{c_{4}(t)}{c_{2}(t)^{2}} + \Or\left(c_{4}(t)^{2}, c_{6}(t),
\ldots \right) \right] .
\end{eqnarray}
Due to the simple scaling $c_{2n}(t) \sim t^{-n (d+3)/4}$ of the
cumulants for large $t$ the expression in the square brackets tends
to a constant. The first correction to the Gaussian
distribution comes from the four-point cumulant $c_{4}(t)$.
The contributions to
$c_{4}(t)$ at second order in $f$ are shown in
figure~\ref{fig5}~(b-d). The first order (figure~\ref{fig5}~(a))
vanishes due to the $x$-derivative associated with the vertex $f$.
This means that upon application of the renormalization group the
amplitude of $c_{4}(t)$ at the fixed point is of the order
$\epsilon^{2}$. Analogously any cumulant $c_{2 n}(t)$ with $n
\geq 2$ is of the order $\epsilon^{n}$ since a non-vanishing
contribution to this function requires at least $n$ vertices
proportional to $f$. Therefore we only need the Gaussian part of
the distribution to calculate $\langle |\psi| \rangle $ at first
order in $\epsilon$. The result is
\begin{equation}
\Phi_{\rm l}(t) = \left( \frac{2}{\pi} \frac{\sqrt{2}
n_{0}}{(8 \pi)^{d/2} \sigma} \right)^{1/2} t^{-(d+3)/8}
\left[ 1 + \frac{\epsilon}{8} \left( 3 \ln 2 - 1 \right)
+ \Or(\epsilon^{2}) \right] . \label{phil}
\end{equation}
Using the inequalities
\begin{equation}
\langle \phi - |\psi| \rangle^{2} \leq \langle (\phi -
|\psi|)^{2} \rangle \leq \langle \phi^{2} \rangle - \langle
\psi^{2} \rangle = \frac{1}{\lambda} \left( -\dot{\Phi}(t)
\right)
\end{equation}
in conjunction with the upper bound $\Phi_{\rm u}(t)$ as
in~\cite{LC1} one can show that the lower bound~(\ref{phil})
gives the exact long-time behaviour of the density, i.e.
$\Phi(t) \simeq \Phi_{\rm l}(t)$ for $t \rightarrow \infty$.

In three dimensions the random velocity field gives a
logarithmic contribution to the density $\Phi(t)$. Using
equation~(\ref{d=3}) one obtains
\begin{equation}
\Phi(t) \simeq \frac{n_{0}^{1/2}}{2 \pi} t^{-3/4} \left(2 f
\ln(\Lambda^{2} t) \right)^{-1/4} .
\end{equation}

\section{Summary and Discussion} \label{concl}

In this paper the effects of a quenched random shear flow
on the long-time behaviour of the A+B-annihilation reaction
have been studied. It is well-known that in dimensions $d<3$
a random velocity field gives rise to enhanced diffusion and
should, therefore, accelerate the reaction process. Similar
to the case of purely diffusive particle motion the system
segregates after a short time into regions of purely A or B
particles. After this initial stage the density decay is
governed by an effective Gaussian distribution for the initial
density fluctuations. 
While a simple model based on anisotropic L\'evy walks
already yields the correct exponent for the density decay one
has to take many-particle correlations generated by the
quenched randomness into account in order to calculate the
amplitude. We have shown how this is possible in the framework
of a systematic expansion in $\epsilon = 3-d$ and computed
the first order term in this expansion.

In the present paper it was assumed that the diffusion
constants $D_{A}$, $D_{B}$ of A and B particles are equal
and that the hopping rate of both species depends in the same
way on $v({\bf y})$. The more general case $D_{A} \neq D_{B}$
with three different disorder couplings $f_{A}$, $f_{B}$,
$f_{AB}$ (instead of $f$) can be treated in a similar way at
least as long as both species are mobile. For unequal
diffusion constants the fields $\phi$ and $\psi$ are coupled
already in the Gaussian part of the action. Using similar
arguments as in reference~\cite{LC1} it can be shown that
this effect changes only the amplitude of $\Phi(t)$.

It would be interesting to check the results of this work
by simulations. In order to observe the asymptotic
long-time behaviour of a disordered system one has to
perform configurational averages unless the system size is
very large. In the case of a random shear flow the
sample-to-sample fluctuations of the density depend on
the linear size $L_{y}$ in the direction perpendicular to
the velocity field. Since a diffusing particle covers a
distance $l_{D}(t) \sim (D t)^{1/2}$ during the time $t$
the density measured at time t represents approximatively
$N \sim L_{y}/l_{D}(t)$ independent configurations of
the flow. Therefore, lack of self-averaging in a finite
sample leads to a relative error of the order
$N^{-1/2} \sim (D t)^{1/4} L_{y}^{-1/2}$. In a pure
system (without shear flow) the asymptotic power law
can be observed at times of the order $D t \sim
10^{5}$~\cite{TW}. The above estimate shows that a
random system with $L_{y} > 1.2 \cdot 10^{5}$ (or
an equivalent number of smaller systems) is required
to reduce the sample-to-sample fluctuations for
$D t \sim 10^{5}$ to less than $5\%$.

\ackn

The author thanks J. Cardy, M. Howard and B. P. Lee for useful discussions.
This work has been supported by grant Oe199/1-1 of the
Deutsche Forschungsgemeinschaft.

\end{document}